\documentclass[pre,twocolumn,9pt]{revtex4}

\newcommand{\dd}{\mathrm{d}}
\newcommand{\ii}{\mathrm{i}}

\usepackage{ae} %%for Computer Modern fonts
\usepackage{amsmath}
\usepackage{amsfonts}
\usepackage{hyperref}
\usepackage{graphicx}
\usepackage[T1]{fontenc}

\begin{document}

\title{``Analytical Continuation'' of Flattened Gaussian Beams}

\author{Riccardo Borghi}
\affiliation{ Dipartimento di Ingegneria Civile, Informatica e delle Tecnologie Aeronautiche, \\
Universit\`{a} ``Roma Tre'', Via Vito Volterra 62, I-00146 Rome, Italy}

%\ociscodes{000.3860, %Mathematical methods in physics 
%260.1960, %Diffraction Theory, 
%350.5500  %Propagation
%}
%\doi{\url{http://dx.doi.org/10.1364/ao.XX.XXXXXX}}

\begin{abstract}
A purely analytical extension of the flattened Gaussian beams [Opt. Commun. \textbf{107,} 335 (1994)] to	 any values of the beam order, is 
here proposed. Due to it, the paraxial propagation problem of axially symmetric, coherent flat-top beams through arbitrary 
$ABCD$ optical systems can  definitely be solved in closed form via a particular bivariate confluent hypergeometric function.
\end{abstract}

\maketitle

\section{Introduction}
\label{Sec:Introduction}

Flat-top  beams  continue to attract a considerable attention in optics:
during the last five years more than sixty papers have been published on the subject.
In order to model flat-top axially symmetric distributions, two classes of different scenarios appeared: 
in the first one, simple analytical profiles were employed, the most known of them being the superGaussian (SG)~\cite{DeSilvestri/Laporta/Magni/Svelto/1988,Parent/Morigne/Lavigne/1992}, 
which is formally defined by
\begin{equation}
\label{Eq:SGProfile}
\begin{array}{l}
\displaystyle
%\mathrm{FG}_N(\xi)\,=\,\exp(-\xi)\,\sum_{m=0}^{N-1}\,\frac 1{m!}\,\xi^{m}\,,
\mathrm{SG}_\nu(\xi)\,=\,\exp(-\xi^{2\nu})\,,
%\,=\,
%\frac 1{N!}\,\Gamma(N+1,\xi^2)\,,
\end{array}
\end{equation}
where $\nu$ denotes a {\em real} parameter which controls the ``flatness'' of the profile, with the particular case $\nu=1$ giving the Gaussian profile.
The symbol $\xi$ denotes a normalized radial transverse position.  
Despite its mathematical simplicity, it  is  well known that Eq.~(\ref{Eq:SGProfile}) does not allow the  
wavefield of paraxially propagated superGaussian (i.e., for  $\nu\ne 1$) beams to be analytically evaluated, even within the simplest scenario, namely, 
free space.

To overcome such difficulty, which two or three decades ago could represent a considerable computational bottleneck in several practical situations,
alternative approaches were proposed in 1994 and in 2002 by Gori and Li, respectively, to conceive analytical models able to solve the free space
propagation problem.
The former was called {\em flattened Gaussian} (FG henceforth)~\cite{Gori/1994}, 
%for which the free-space paraxial propagation was studied in exact terms~\cite{Bagini/Borghi/Gori/Pacileo/Santarsiero/Ambrosini/Schirripa/1996}. 
and, different from SG,  is expressed through an explicit {finite sum} of terms, namely 
\begin{equation}
\label{Eq:FGProfile}
\begin{array}{l}
\displaystyle
%\mathrm{FG}_N(\xi)\,=\,\exp(-\xi)\,\sum_{m=0}^{N-1}\,\frac 1{m!}\,\xi^{m}\,,
\mathrm{FG}_N(\xi)\,=\,\exp(-N\xi^2)\,\sum_{m=0}^{N-1}\,\frac {(N\xi^{2})^{m}}{m!}\,,
%\,=\,
%\frac 1{N!}\,\Gamma(N+1,\xi^2)\,,
\end{array}
\end{equation}
%
%where $\Gamma(\cdot,\cdot)$ denotes the incomplete gamma function~\cite{DLMF}.
where the {integer} parameter $N>0$ will be referred to as the FG order. 
Scaling the $\xi$ variable by the factor $\sqrt{N}$ gives the FG transverse profile a flat-topped shape which, for $N=1$, reduces to a 
Gaussian distribution,  whereas for $N\to\infty$ tends to the characteristic function of the unitary disk~\cite{FootNote}. The model is computationally exact,  
since the initial distribution~[Eq.~(\ref{Eq:FGProfile})] can be recast in terms of a superposition of $N$ {\em standard } Laguerre-Gauss (sLG henceforth) beams. 
Accordingly, in order to evaluate the field propagated in free space, it  is sufficient to sum up the $N$ propagated sLG, a job which can exactly be done, 
always~\cite{Bagini/Borghi/Gori/Pacileo/Santarsiero/Ambrosini/Schirripa/1996}.
In~\cite{Borghi/2001},  a different superposition scheme of the profile~[Eq.~(\ref{Eq:FGProfile})] was proposed, in which the sLG family was replaced by the 
so-called {\em elegant} Laguerre-Gauss (eLG henceforth) set. 
In this way, not only free-space propagation, but also the interaction of FG beams with {\em any} axially symmetric paraxial optical system can be dealt with 
in exact terms, always through finite sums.   

In 2002,  Yaijun Li proposed an analytical model alternative to the FG one. %To this end,  
The idea was to impose a local ``flatness'' condition, which required the first  $N-1$ even $\xi$-derivatives of the profile to be null at the origin $\xi=0$~\cite{Li/2002}. 
%On using such condition, %which is satisfied by the FG profile in Eq.~(\ref{Eq:FGProfile}), 
More precisely,  Li conceived the following analytical model:
\begin{equation}
\label{Eq:LiProfile}
\begin{array}{l}
\displaystyle
%\mathrm{FG}_N(\xi)\,=\,\exp(-\xi)\,\sum_{m=0}^{N-1}\,\frac 1{m!}\,\xi^{m}\,,
\mathrm{LiG}_N(\xi)\,=\,\sum_{m=1}^{N}\,(-1)^{m-1}\,\left({{N}\atop{m}}\right)\,\exp(-m \xi^2)\,=\,\\
\\
\,=\,\dfrac{\left\{1-\left[1-\exp\left(-\xi^2\right)\right]^N\right\}}N\,,
%\,=\,
%\frac 1{N!}\,\Gamma(N+1,\xi^2)\,,
\end{array}
\end{equation}
which, different from FG, is  based on the superposition of $N$ fundamental Gaussian beams having variable widths. 
In particular, it is not difficult to prove that~\cite{Li/2002}
\begin{equation}
\label{Eq:LiFlatness.1}
\begin{array}{l}
\displaystyle
\left[\dfrac{\dd^{2k}}{\dd \xi^{2k}}\mathrm{LiG}_N(\xi)\right]_{\xi=0}\,=\,0\,,\qquad\qquad 1 \le k < N\,,
\end{array}
\end{equation}
the odd derivatives being identically null, due to the radial symmetry of the function $\mathrm{LiG}_N(\xi)$.

Both Gori's and Li's models provide exact solutions to the paraxial propagation problem of coherent, axially symmetric flat-top beams. 
From a merely mathematical perspective, their only own limit is represented by the fact that, different from SG, only positive integer orders $N$ 
can be dealt with to describe the initial flat-top distribution.
It is important to  mention that, for 1D geometry (or rectangular 2D geometries), general analytical 
solutions were already provided, at least upon free propagation, by modeling the flat-top profile  
via an error function~\cite{Sedukhin/2015}. 
An attempt to extend the 2D circular FG model to noninteger orders was also proposed 
in~\cite{Borghi/2013}, but only approximate estimates of the 
free-space propagated field were obtained within the asymptotic limit $N\gg 1$.
 
The aim of the present paper is to solve exactly the propagation problem of FG beams of {\em any} order (real or even complex) through 
typical axially symmetric paraxial optical systems. To this end, the right side of Eq.~(\ref{Eq:FGProfile}) will first be identified as an incomplete Gamma 
functions, which is known to be defined onto the whole complex plane, as far as both arguments are 
concerned. 
An immediate byproduct of such identification will be the closed form expression of the $M^2$ factor of 
FG beams of any order, an interesting generalization of the result found in~\cite{Bagini/Borghi/Gori/Pacileo/Santarsiero/Ambrosini/Schirripa/1996}. 
This is shown in Sec.~\ref{Sec:FGModel.1} of the present paper. 
The most important results are presented in Secs.~\ref{Sec:FreeSpaceParaxialPropagation} and~\ref{Sec:ABCDParaxialPropagation}.
 In the former, the free-space propagation problem will  be solved thanks to an important class of integrals recently evaluated by Yuri Brychkov.
 Although the more general propagation problem will be solved in Sec.~\ref{Sec:ABCDParaxialPropagation}, the analysis presented in 
 Sec.~\ref{Sec:FreeSpaceParaxialPropagation} should be viewed as an important propaedeutical step. There, it will be shown 
 that a very important, but nevertheless not so much known, class of special functions, called {\em bivariate 
hypergeometric functions}, together with the corresponding confluent versions, forms the mathematical skeleton of the paraxially diffracted wavefield. Bivariate 
hypergeometric were first introduced in~1880 by 
Paul Appell~\cite{Appell/1880}, their confluent version forty years later by Paul Humbert~\cite{Humbert/1922}.
The results we are going to present would also give readers a partial answer about the lack, for more than thirty years, of purely analytical solutions to the 
problem of the paraxial propagation of
coherent 2D flat-topped beams. 

The present work has a clear mathematical character: for instance, dimensionless quantities will be used wherever possible. 
Moreover, the number of mathematical appendices have been mostly limited, because we strongly believe that following all important 
mathematical steps could greatly help readers to fully grasp the essence of our analysis, as well as the importance of such still mysterious special functions, which  will lead to analytical, elegant, and exact solutions.

\section{Preliminaries}
\label{Sec:FGModel.1}

\subsection{``Analytical continuation'' of the FG model}
\label{Subsec:Continuation}

Already in 1996, Sheppard \& Saghafi~\cite{Sheppard/Saghafi/1996} pointed out that  Eq.~(\ref{Eq:FGProfile}) can be given the following closed form:
\begin{equation}
\label{Eq:FGProfileGamma}
\begin{array}{l}
\displaystyle
\mathrm{FG}_N(\xi)\,=\,%\exp(-\xi^2)\,\sum_{m=0}^N\,\frac 1{m!}\,\xi^{2m}\,,
%\,=\,
\dfrac {\Gamma(N,\,N\xi^2)}{\Gamma(N)}\,,
\end{array}
\end{equation}
where $\Gamma(\cdot)$ and $\Gamma(\cdot,\cdot)$ denote gamma and incomplete gamma functions, respectively~\cite{DLMF}.
Different from Eq.~(\ref{Eq:FGProfile}),  Eq.~(\ref{Eq:FGProfileGamma}) is not limited to integer FG orders, but rather it can be analytically continued to real 
and also complex values of $N$. 

\begin{figure}[!ht]
\centering
\begin{minipage}[t]{4.5cm}
\centerline{\includegraphics[width=4.5cm,angle=-0]{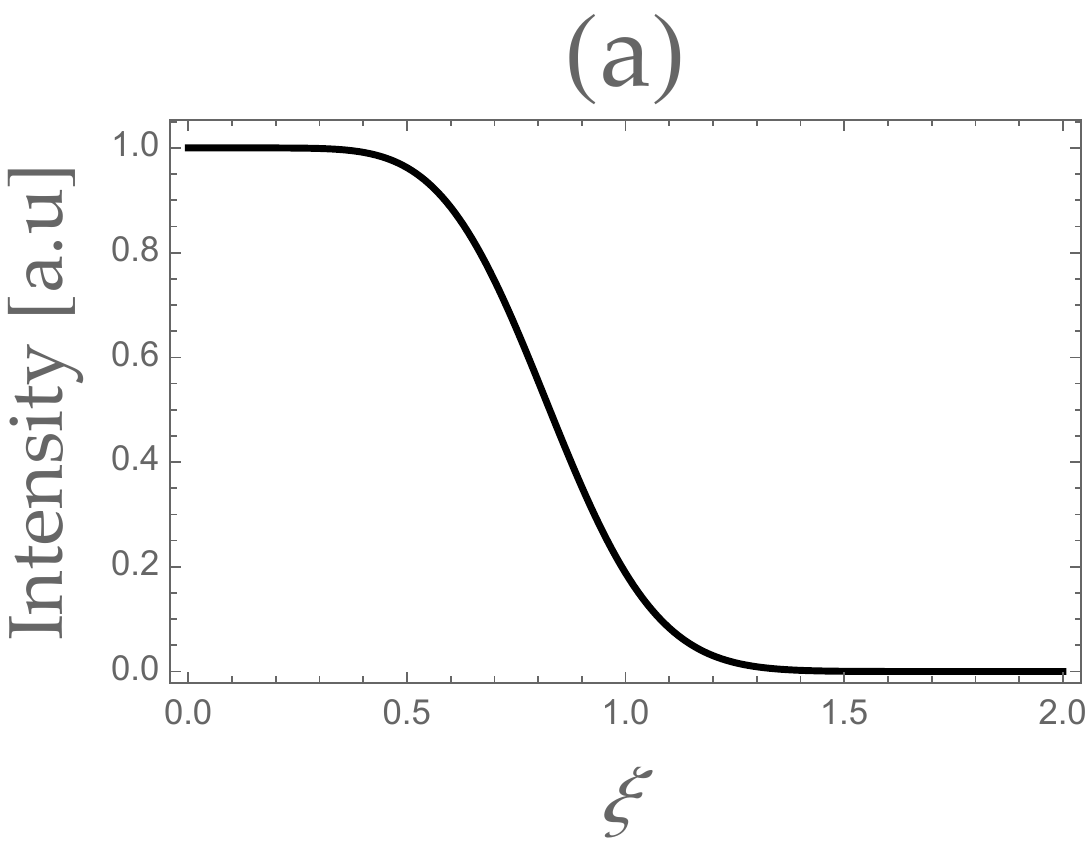}}
\centerline{\includegraphics[width=4.5cm,angle=-0]{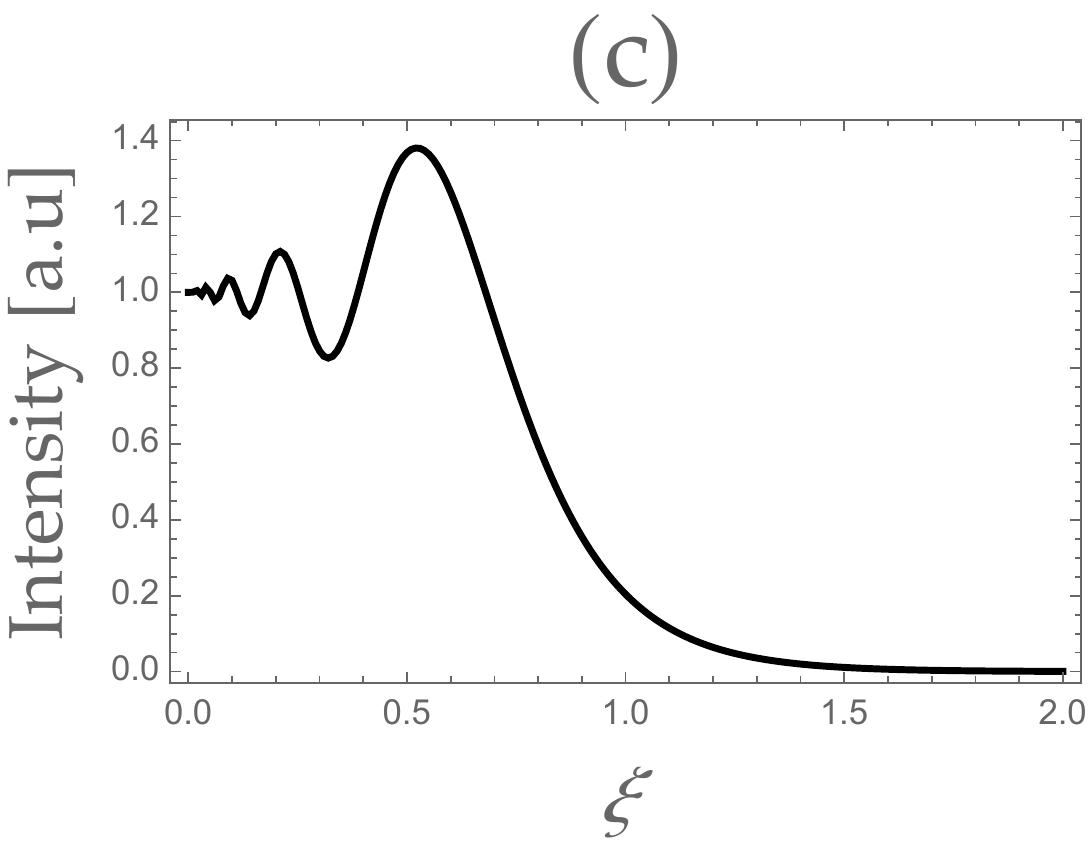}}
\end{minipage}
\hspace*{1mm}
\begin{minipage}[t]{4.5cm}
\centerline{\includegraphics[width=4.5cm,angle=-0]{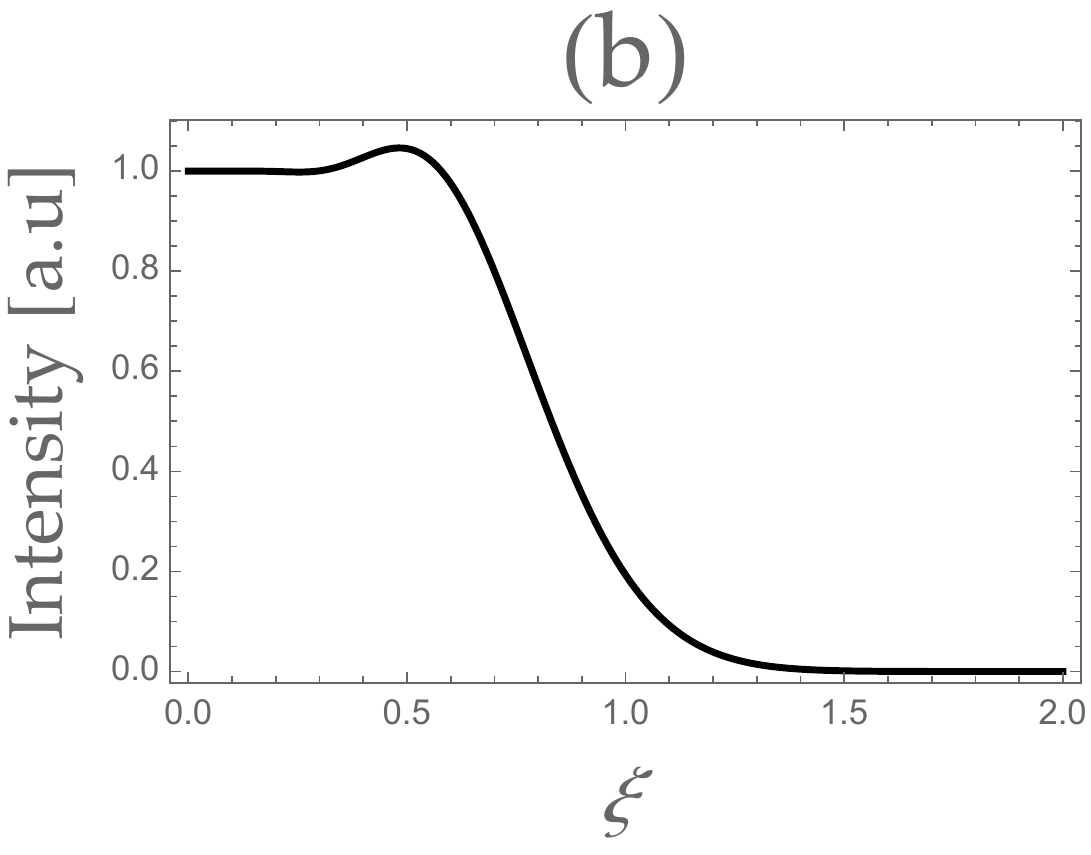}}
\centerline{\includegraphics[width=4.5cm,angle=-0]{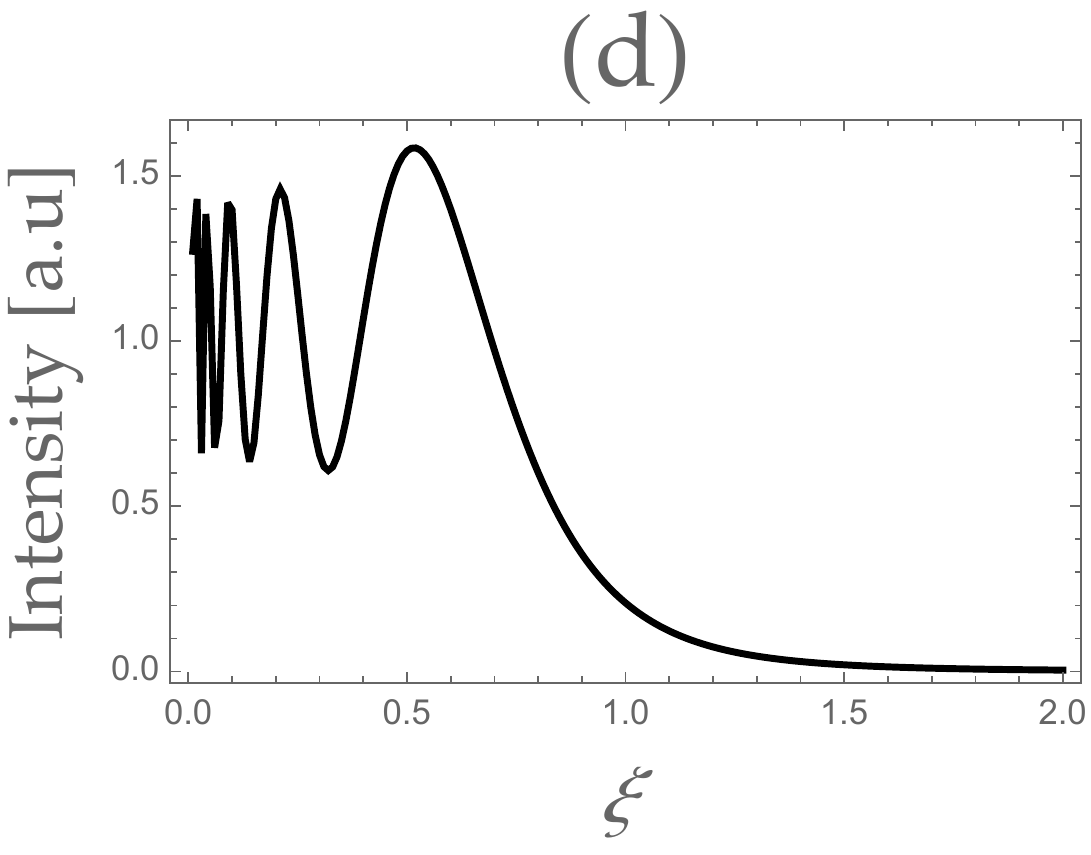}}
\end{minipage}
\caption{ Behaviour of the intensity profile $|\mathrm{FG}_N(\xi)|^2$ for complex 
orders~$N=4~\exp(\ii \phi)$, with $\phi=0$ (a), $\phi=\pi/4$ (b), $\phi=9\pi/20$ (c), and $\phi=\pi/2$ (d).
Figures (a) and (b) correspond to $\mathrm{Re}\{N\}>1$, the opposite for figures (c) and (d).}
\label{Fig:FGModel.1}
\end{figure}

As a preliminary result of the extended definition into Eq.~(\ref{Eq:FGProfileGamma}), an analytical check of  Li's ``flatness condition'' described in the previous section %~\cite{Li/2002}
will now be carried out.
To this end, it is not difficult to prove, on using  formulas~1.1.1.1 and~1.8.1.17 of~\cite{Brychkov/2008}, together with long but simple algebra, that
\begin{equation}
\label{Eq:DerivativeGamma}
\begin{array}{l}
\displaystyle
\dfrac{\dd^n}{\dd \xi^n}\Gamma(N,N\xi^2)\,=\,
-2^nn! N^N\,\exp(-N\xi^2)\,\xi^{2N-n}\,\\
\\
\displaystyle
\times\sum_{k=0}^{[n/2]}\,
\dfrac{(n-k-1)!}{4^k k! (n-2k)!}\,L^{(N-n+k)}_{n-k-1}(N\xi^2)\,,\qquad\qquad n\,\ge\,1\,.
\end{array}
\end{equation}
Then, on taking the axial symmetry of the $\mathrm{FG}_N(\xi)$ distribution into account, from Eq.~(\ref{Eq:DerivativeGamma}), it follows, similar to Eq.~(\ref{Eq:LiFlatness.1})
for Li's model, that
\begin{equation}
\label{Eq:FGProfileGamma.2}
\begin{array}{l}
\displaystyle
\left[\dfrac{\dd^n}{\dd \xi^n}\Gamma(N,N\xi^2)\right]_{\xi=0}\,=\,0\,,\qquad\qquad \left\{{{2 \le n < 2\,\mathrm{Re}\{N\}}\atop{n\,\,\,\mathrm{even}}}\right\}
\end{array}
\end{equation}
the odd derivatives being also identically null. Equation~(\ref{Eq:FGProfileGamma.2}) implies the real part of $N$ to be chosen greater than 1.
For real values of FG's order, it is known that such a limitation guarantees the absence of a cusp at the 
origin $\xi=0$. % (the second derivative would be infinite). %(the limiting case of $N=1$, i.e., the Gaussian profile, would be at the border). 
If complex values of $N$ were involved, order's imaginary part would be responsible for extra oscillations of the intensity profile. Just to give a single visual example, 
in Fig.~\ref{Fig:FGModel.1} 
the behaviour of the intensity $|\mathrm{FG}_N(\xi)|^2$ is plotted for $N=4 \exp(\ii \phi)$, with $\phi=0$ (a), $\phi=\pi/4$ (b), $\phi=9\pi/20$ (c), and $\phi=\pi/2$ (d).
Notice that figures (a) and (d) correspond to the extreme cases of imaginary part and real part of $N$ null, respectively.

\subsection{Spreading properties: closed form expression of the $M^2$ factor}
\label{Subsec:SpreadingProperties}

An interesting byproduct of the $\Gamma$-based definition into Eq.~(\ref{Eq:FGProfileGamma}) is the extension of the $M^2$ factor evaluation of FG 
beams (first established in~\cite{Bagini/Borghi/Gori/Pacileo/Santarsiero/Ambrosini/Schirripa/1996} for $N\in\mathbb{N}$) also to noninteger orders.
To this end, consider an initial field distribution across the plane $z=0$ of a cylindrical reference frame $(\boldsymbol{r},z)$, say $\psi_0(\boldsymbol{r})$, 
given by
\begin{equation}
\label{Eq:FlattenedGaussian.0}
\begin{array}{l}
\displaystyle
\psi_0(\boldsymbol{r})\,=\,\mathrm{FG}_N\left(\dfrac {r}{a}\right)\,=\,\dfrac {\Gamma\left(N,N \dfrac{r^2}{a^2}\right)}{\Gamma(N)}\,,
\end{array}
\end{equation}
where an overall amplitude constant has been set to one  and the symbol $a$ denotes the ``width'' of the flat top field distribution.
For simplicity, it will be set $a=1$. 

The evaluation of the $M^2$ factor, which is defined as the product of the normalized second order moments across the $z=0$ and the spatial frequency
planes is detailed  in~Appendix~\ref{App:M2}, where it is proved the following closed-form expression: 
\begin{equation}
\label{Eq.M2FactorFG}
\begin{array}{l}
\displaystyle
M_{\rm FG}^2\,=\,%2\pi\,\sigma_r\,\sigma_p\,=\,
\dfrac{\sqrt{(N+1)\dfrac{\Gamma(N+1/2)}{\sqrt\pi\,\Gamma(N+1)}\left[1\,-\,\dfrac{\Gamma(N+3/2)}{\sqrt\pi\,\Gamma(N+2)}\right]}}{1\,-\,\dfrac{\Gamma(N+1/2)}{\sqrt\pi\,\Gamma(N+1)}}\,,
\end{array}
\end{equation}
which generalizes the 1996 analysis of~\cite{Bagini/Borghi/Gori/Pacileo/Santarsiero/Ambrosini/Schirripa/1996} to $N\notin\mathbb{N}$. It is worth comparing Eq.~(\ref{Eq.M2FactorFG}) with the corresponding expression
of SG beam $M^2$  factor, namely~\cite{Parent/Morigne/Lavigne/1992}
\begin{equation}
\label{Eq.M2FactorSG}
\begin{array}{l}
\displaystyle
M_{\rm SG}^2\,=\,\dfrac{\sqrt{\Gamma(2/\nu)}}{\Gamma(1/\nu)/\nu}\,,
\end{array}
\end{equation}
deceptively simpler. 
{The mathematical elegance of Eq.~(\ref{Eq.M2FactorFG}) also reveals some practical usefulness. 
Suppose to be interested in solving the following problem: given a SG beam (i.e., given $\nu$), what is the FG beam (i.e., the value of $N$) having identical spreading properties (i.e., the same $M^2$ factor)?
A partial answer to such problem was already been provided into Ref.~\cite{Bagini/Borghi/Gori/Pacileo/Santarsiero/Ambrosini/Schirripa/1996}, where  it was proved that $M_{\rm FG}^2$ can be estimated, 
within the asymptotic limit $N\gg 1$, by $(N/\pi)^{1/4}$, so that 
\begin{equation}
\label{Eq.M2FactorSG.00}
\begin{array}{l}
\displaystyle
N\,\sim\,\pi\,\left[M_{\rm SG}^2(\nu)\right]^4\,=\,\pi\,\nu^4\,\left[\dfrac{\Gamma(2/\nu)}{\Gamma^2(1/\nu)}\right]^2\,,\qquad N,\nu \gg 1\,.
\end{array}
\end{equation}
The correspondence between $\nu$ and $N$ given into Eq.~(\ref{Eq.M2FactorSG}) can be considerably improved, especially within the value range close to the unity, by 
asymptotically expanding the right side of Eq.~(\ref{Eq.M2FactorFG}) as follows:
\begin{equation}
\label{Eq.M2FactorFGAsymptotics}
\begin{array}{l}
\displaystyle
M_{\rm FG}^2\,=\,
\left(\dfrac N\pi\right)^{1/4}\,
\left[
1\,+\,
\dfrac{7\pi\,-\,8}{16\,\pi^2}\,\dfrac \pi N\,+\,\mathcal{O}\left(\dfrac 1N\right)
\right]\,,
\end{array}
\end{equation}
a result which can be achieved by using, for instance, the Mathematica command \texttt{Series}. Then, on again imposing $M_{\rm FG}^2\,=\,M_{\rm SG}^2$ and after taking Eq.~(\ref{Eq.M2FactorFGAsymptotics}) into account, it is not difficult to prove that the quantity $X\,=\,(N/\pi)^{1/4}$ must satisfy the following fourth-order algebraic equation:
\begin{equation}
\label{Eq.M2FactorFGAsymptotics.2}
\begin{array}{l}
\displaystyle
X^4\,-\,M_{\rm SG}^2(\nu)\,X^3\,+\,\dfrac{7\pi\,-\,8}{16\,\pi^2}\,=\,0\,.
\end{array}
\end{equation}
It is possible to express $X$ in closed form via Cardano's formula. For reader's convenience, the analytical expression has been reported into Appendix~\ref{App:Cardano}. 
Just to give a visual idea of the degree of approximation of the asymptotics in Eq.~(\ref{Eq.M2FactorFGAsymptotics}), in Fig.~\ref{Fig:ComparisonM2FGSG} the behaviour of 
$M_{\rm SG}^2$ is plotted against $\nu$ according to Eq.~(\ref{Eq.M2FactorSG}) (solid curve). In the same figure, also the behaviour of
the quality factor of FG beams given by Eq.~(\ref{Eq.M2FactorFG}) is shown
when the FG order $N$ is chosen according to the simplest choice given into Eq.~(\ref{Eq.M2FactorSG.00}) (dashed curve) as well as that improved through 
the algorithm described by Eq.~(\ref{Eq.M2FactorFGAsymptotics.2})  (dotted curve).
\begin{figure}[!ht]
\centering
\centerline{\includegraphics[width=8cm,angle=-0]{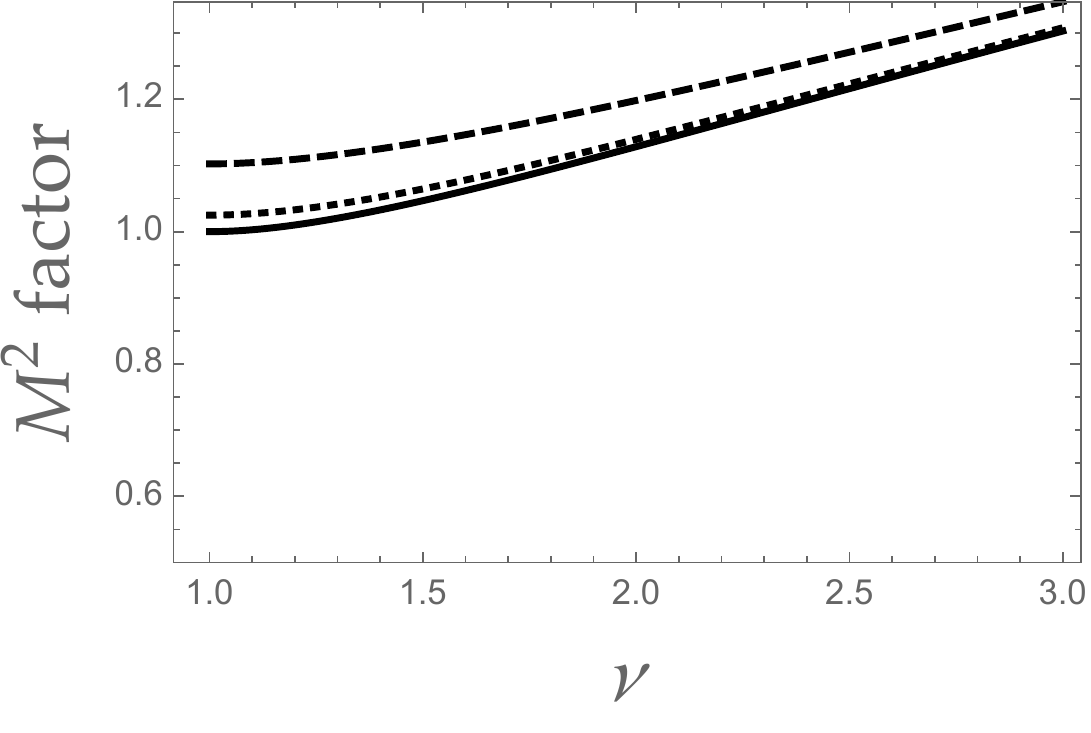}}
\caption{ Behaviour of $M_{\rm SG}^2$ against $\nu$ according to Eq.~(\ref{Eq.M2FactorSG}) (solid curve), together with
the behaviour of $M_{\rm FG}^2$ given by Eq.~(\ref{Eq.M2FactorFG}) and evaluated for FG orders $N$ chosen according to 
Eq.~(\ref{Eq.M2FactorSG.00}) (dashed curve) as well as  through the algorithm described by Eq.~(\ref{Eq.M2FactorFGAsymptotics.2})
and Appendix~\ref{App:Cardano} (dotted curve).}
\label{Fig:ComparisonM2FGSG}
\end{figure}
}

In the next two sections, our extension of the FG model will further reveal its 
powerfulness and mathematical elegance. 

\section{Free-space paraxial propagation of FG beams}
\label{Sec:FreeSpaceParaxialPropagation}

\subsection{Preliminaries}
\label{SubSec:FreeSpaceParaxialPropagationPreliminaries}

Suppose the initial field distribution given by Eq.~(\ref{Eq:FlattenedGaussian.0}) is allowed to propagate in free space. 
The corresponding field, say $\psi(\boldsymbol{r};z)$, can be expressed, apart from an overall phase factor $\exp(\ii kz)$, as follows:
\begin{equation}
\label{Eq:FlattenedGaussian.0.1}
\begin{array}{l}
\displaystyle
\psi(\boldsymbol{r};z)\,=\,-\dfrac{\ii\,U}{2\pi}\,\int_{\mathbb{R}^2}\,\dd^2\rho\,\psi_0(\boldsymbol{\rho})\,\exp\left(\dfrac{\ii U}2\,\left|\boldsymbol{r}\,-\,\boldsymbol{\rho}\right|^2\right)\,,
\end{array}
\end{equation}
where the Fresnel number $U=ka^2/z$ has been introduced and the beam width $a$ has been used as unit for measuring all transverse sizes. 
This means that the quantity $\boldsymbol{r}$ should be meant as the ratio between the transverse position vector of the observation point and $a$. 
For integer FG orders, the free space propagation problem has already been solved in~\cite{Gori/1994} by expanding the initial field distribution $\psi_0$ as 
the linear combination of a finite number of sLG beams. 
It is then sufficient to propagate each sLG beam up to the observation plane and to recombine all of them with the initial expanding 
coefficients for the correct value of $\psi(\boldsymbol{r};z)$ to be retrieved. 
As we are going to show in a moment, the $\Gamma$-based model into Eq.~(\ref{Eq:FGProfileGamma}) allows an exact evaluation of the propagated 
wavefield~(\ref{Eq:FlattenedGaussian.0.1}) also for $N\notin \mathbb{N}$. 
It is worth recalling that, from a mere practical perspective, the present section could seem somewhat 
redundant, as in Sec.~\ref{Sec:ABCDParaxialPropagation} the more general propagation problem within $ABCD$ systems will be solved. 
Nevertheless, we believe what is contained in the present section could help nonspecialist readers to familiarize with the main notations and 
mathematical tools which will constitute the basis of the general results presented into Sec.~\ref{Sec:ABCDParaxialPropagation}. 
In other words, it should be considered as a useful, propaedeutical material.

We start on substituting from Eqs.~(\ref{Eq:FlattenedGaussian.0}) into Eq.~(\ref{Eq:FlattenedGaussian.0.1}), which after simple algebra gives
\begin{equation}
\label{Eq:FlattenedGaussian.0.1.1}
\begin{array}{l}
\displaystyle
\psi(\boldsymbol{r};z)\,=\,-\dfrac{\ii\,U}{\Gamma(N)}\,\exp\left(\dfrac{\ii U\,r^2}2\right)\,\\
\\
\displaystyle
\times\,
\int_0^\infty\,\dd\rho\,\rho\,\exp\left(\dfrac{\ii U}2\,\rho^2\right)\,{\Gamma\left(N, N\,\rho^2\right)}\,J_0(U r\,\rho)\,,
\end{array}
\end{equation}
where $J_0$ denotes the 0th-order Bessel function of the first kind.
It is worth  recasting the incomplete $\Gamma$ function as 
\begin{equation}
\label{Eq:FlattenedGaussian.0.1.1Modified.2}
\begin{array}{l}
\displaystyle
\dfrac{\Gamma(N, N\xi)}{\Gamma(N)}\,=\,1\,-\,\dfrac{\gamma(N, N\xi)}{\Gamma(N)}\,,
\end{array}
\end{equation}
where $\gamma(\cdot,\cdot)$ denotes the ``lower'' incomplete gamma function. Then Eq.~(\ref{Eq:FlattenedGaussian.0.1.1}) takes on the form
\begin{equation}
\label{Eq:FlattenedGaussian.0.1.1Modified.3}
\begin{array}{l}
\displaystyle
\psi(\boldsymbol{r};z)\,=\,\\
\\
\displaystyle
\,=\,-{\ii\,U\,\exp\left(\dfrac{\ii U\,r^2}2\right)}
\,\int_0^\infty\,\dd\rho\,\rho\,\exp\left(-\dfrac U{2\ii}\,\rho^2\right)\,J_0(U r\,\rho)\,\\
\\
\displaystyle
\,+\,\dfrac{\ii\,U\,\exp\left(\dfrac{\ii U\,r^2}2\right)}{\Gamma(N)}\,\\
\\
\displaystyle
\qquad\,\times\int_0^\infty\,\dd\rho\,\rho\,\exp\left(-\dfrac U{2\ii}\,\rho^2\right)\,\gamma\left(N, N\rho^2\right)\,J_0(U r\,\rho)\,.
\end{array}
\end{equation}
The first term is identically equal to one 
(it is nothing but a unitary plane wave propagating along the $z$-axis). As far as the second is concerned, the following notable formula has recently been 
published by  Brychkov~\cite[formula~9.2.20]{Brychkov/2014}: %~({\bf CHECKED 23 nov 2022}):
\begin{equation}
\label{Eq:BrychkovNewFormula}
\begin{array}{l}
\displaystyle
\int_0^\infty\,\dd x\,
x^{\alpha-1}\,\exp(-a\,x^2)\,\gamma(\mu,b x^2)\,J_\nu(c\,x)\,=\,\\
\\
\,=\,
\dfrac{2^{-\nu-1}b^\mu c^\nu \Gamma\left(\mu+\dfrac{\alpha+\nu}2\right)}{\mu a^{\mu+(\alpha+\nu)/2}\Gamma(\nu+1)}\,
\Psi_1\left(
\left.
\begin{array}{l}
\mu+\dfrac{\alpha+\nu}2,\mu\\
\mu+1,\nu+1
\end{array}
\right| -\dfrac ba, -\dfrac{c^2}{4a}
\right)\,.
\end{array}
\end{equation}
%
%under the conditions
%%
%\begin{equation}
%\label{Eq:BrychkovNewFormula.2}
%\begin{array}{l}
%\displaystyle
%\mathrm{Re}\{\alpha+2\mu+\nu\}>0\,,\,\mathrm{Re}\{a\}>0\,,\,\mathrm{Re}\{a+b\}>0\,.
%\end{array}
%\end{equation}
%%
Then, on using Eqs.~(\ref{Eq:FlattenedGaussian.0.1.1Modified.3}) and~(\ref{Eq:BrychkovNewFormula}), long but straightforward algebra gives
\begin{equation}
\label{Eq:FlattenedGaussian.0.1.1Modified.4}
\begin{array}{l}
\displaystyle
\psi(\boldsymbol{r};z)\,=\,
1\,-\,\exp\left(\dfrac{\ii U\,r^2}2\right)\,\left(\dfrac{2\ii N}U\right)^{N}\,\\
\\
\times\,
\Psi_1\left(
\left.
\begin{array}{l}
N+1,N\\
N+1,1
\end{array}
\right| -\dfrac{2\ii N}U, -\dfrac{\ii U\,r^2}2
\right)\,.
\end{array}
\end{equation}

\subsection{Short Tour on Bivariate Hypergeometric Functions}
\label{SubSec:ShortTourBivariateHypergeometricFunctions}

The symbol  $\Psi_1$ into Eq.~(\ref{Eq:FlattenedGaussian.0.1.1Modified.4}) denotes a special function called {\em bivariate confluent hypergeometric}.
It is worth briefly describing the principal definitions and properties which are important for our scopes. 
Function $\Psi_1$ is formally defined through the following double series power expansion:
\begin{equation}
\label{Eq:Psi1}
\begin{array}{l}
\displaystyle
\Psi_1\left(
\left.
\begin{array}{l}
a,b\\
c,c'
\end{array}
\right| z, w
\right)\,=\,
\sum_{k=0}^\infty\,\sum_{\ell=0}^\infty\,\dfrac{(a)_{k+\ell}\,(b)_k}{(c)_k(c')_\ell}\,\dfrac{z^k}{k!}\,\dfrac{w^\ell}{\ell!}\,,%\qquad\quad |z| \le 1\,, |w| \le 1\,,
\end{array}
\end{equation}
valid for $|z|\le 1$. The symbol $(\cdot)_n$ denotes Pochhammer's symbol. 
Another bivariate confluent hypergeometric function which will be met in the present paper is the function $\Phi_1$, defined by 
\begin{equation}
\label{Eq:Phi1}
\begin{array}{l}
\displaystyle
\Phi_1\left(
\left.
\begin{array}{c}
a,b\\
c
\end{array}
\right| z, w
\right)\,=\,
\sum_{k=0}^\infty\,\sum_{\ell=0}^\infty\,\dfrac{(a)_{k+\ell}\,(b)_k}{(c)_{k+\ell}}\,\dfrac{z^k}{k!}\,\dfrac{w^\ell}{\ell!}\,,%\qquad\quad |z| \le 1\,,|w| \le 1\,.
\end{array}
\end{equation}
valid for $|z|\le 1$.
Functions $\Psi_1$ and $\Phi_1$ are members of a family of functions that generalize Kummer's confluent hypergeometric function ${}_1F_1$.
In particular,  $\Phi_1$ is obtained from the so-called Appell function $F_1$, defined by
\begin{equation}
\label{Eq:F1}
\begin{array}{l}
\displaystyle
F_1\left(
\left.
\begin{array}{c}
a,b_1,b_2\\
c
\end{array}
\right| z, w
\right)\,=\,
\sum_{k=0}^\infty\,\sum_{\ell=0}^\infty\,\dfrac{(a)_{k+\ell}\,(b_1)_k\,(b_2)_\ell}{(c)_{k+\ell}}\,\dfrac{z^k}{k!}\,\dfrac{w^\ell}{\ell!}\,,%\qquad\quad |z| \le 1\,,|w| \le 1\,,
\end{array}
\end{equation}
(again valid for $|z|\le 1$), through the following limiting definition:
\begin{equation}
\label{Eq:Phi1F1}
\begin{array}{l}
\displaystyle
\Phi_1\left(
\left.
\begin{array}{c}
a,b\\
c
\end{array}
\right| z, w
\right)\,=\,
\lim_{\epsilon\to 0}\,
F_1\left(
\left.
\begin{array}{c}
a,b,\dfrac 1\epsilon\\
c
\end{array}
\right| z, \epsilon w
\right)\,,
\end{array}
\end{equation}
which can be proved on first substituting the  identity
\begin{equation}
\label{Eq:Confluenza}
\begin{array}{l}
\displaystyle
\lim_{\epsilon\to 0}\,
\left(\dfrac 1\epsilon\right)_\ell\,\epsilon^\ell\,=\,1\,,
%\\
%\\
%\,=\,
%\displaystyle
%\lim_{\epsilon\to 0}\,
%\epsilon^\ell\,\dfrac 1\epsilon\,\left(\dfrac 1\epsilon\,+\,1\right)\,\left(\dfrac 1\epsilon\,+\,2\right)\,\ldots\,\left(\dfrac 1\epsilon\,+\,\ell\,-\,1\right)\,=\,\\
%\\
%\,=\,
%%\epsilon^\ell\,\dfrac 1{\epsilon^\ell}\,
%\displaystyle
%\lim_{\epsilon\to 0}\,(1\,+\epsilon)\,(1\,+2\epsilon)\,\ldots\,[1\,+(\ell-1)\epsilon]\,=\,1\,,
\end{array}
\end{equation}
directly into Eq.~(\ref{Eq:Phi1F1}), then on interchanging the limit with the double series.

Multivariate hypergeometric and confluent hypergeometric functions play a role of considerable importance in theoretical physics and applied math. 
In optics, the role of bivariate confluent hypergeometric functions in describing a large class of paraxial optical disturbances has recently been 
pointed out~\cite{ElHalba/Nebdi/Boustimi/Belafhal/2014,Belafhal/Saad/2017}.
Moreover, it is worth stressing that, from a practical viewpoint, Appell's function $F_1$ %defined into Eq.~(\ref{Eq:Phi1.2.1})
is nowadays part of the symbolic platform {\em Mathematica}, %(releases 12 and 13), 
where it is computable with arbitrarily high accuracies.  Also the whole family of  Appell functions, including $F_1$ as well as its three sisters 
$F_2$, $F_3$, and $F_4$, are currently implemented in the latest release of {\em Maple}. It is then highly desirable that in a near future also the set of 
bivariate confluent hypergeometric functions, including $\Psi_1$ and $\Phi_1$, could become part of such a family of ``evaluable'' special functions. In the meanwhile, someone might rightly object to the practical usefulness of functions that are defined through
double infinite series like those into Eqs.~(\ref{Eq:Psi1})\,-\,(\ref{Eq:F1}). 
To overcome such difficulties, some tricks will be implemented in the rest of the paper, tricks aimed at extending the validity domain 
of $\Psi_1$ and $\Phi_1$ beyond the series definitions, and then to improve the practical usefulness of our analytical results.

Function $\Psi_1$ can be continued  by using the following  transformation~\cite[formula~2.54]{Choi/Hasanov/2011}:
\begin{equation}
\label{Eq:RiduzionePsi1}
\begin{array}{l}
\displaystyle
\Psi_1\left(
\left.
\begin{array}{l}
\alpha,\beta\\
\gamma_1,\gamma_2
\end{array}
\right| z, w
\right)\,=\,	\\
\\
\,=\,
\dfrac 1{(1-z)^\alpha}\,
\Psi_1\left(
\left.
\begin{array}{l}
\alpha,\gamma_1-\beta\\
\gamma_1,\gamma_2
\end{array}
\right| \dfrac z{z-1}, \dfrac w{1-z}
\right)\,,
\end{array}
\end{equation}
which, once substituted into Eq.~(\ref{Eq:FlattenedGaussian.0.1.1Modified.4}), gives a new, closed-form, expression of the paraxial propagated field
\begin{equation}
\label{Eq:PropagatedFieldNew}
\begin{array}{l}
\displaystyle
%\boxed{
\psi(\boldsymbol{r};z)\,=\,
1\,-\,\dfrac{\exp\left(\dfrac{\ii U\,r^2}2\right)}{1\,+\,\dfrac{2\ii N}U}\,\left(\dfrac 1{1\,+\,\dfrac U{2\ii N}}\right)^{N}\,\\
\\
\times\,
\Psi_1\left(
\left.
\begin{array}{l}
N+1,1\\
N+1,1
\end{array}
\right| \dfrac 1{1\,+\,\dfrac U{2\ii N}}, -\dfrac{\dfrac{\ii U\,r^2}2}{1\,+\,\dfrac{2\ii N}U}
\right)\,,
%}
\end{array}
\end{equation}
indubitably one of the main results of the present paper. 

Waiting for {\em Mathematica} or {\em Maple} to develop their own built-in version of $\Psi_1$, it is worth working on the expression 
into Eq.~(\ref{Eq:PropagatedFieldNew}) by using a notable integral representation found again in~\cite{Choi/Hasanov/2011}.
For the sake of clarity, all mathematical steps are confined into~Appendix~\ref{App:Choi/Hasanov}, where it is proved that
\begin{equation}
\label{Eq:IntegralRepresentationPsi1.7}
\begin{array}{l}
\displaystyle
%\boxed{
\Psi_1\left(
\left.
\begin{array}{l}
N+1,1\\
N+1,1
\end{array}
\right| x, y
\right)
\displaystyle
\,=\,\\
\\
\,=\,
\displaystyle
N\,\int_0^1\dd\xi\,
\dfrac{(1-\xi)^{N-1}}{(1-x\xi)^{N+1}}\,
{}_1F_1\left(N+1;1;\dfrac {y}{1-x\xi}\right)\,.
%}
\end{array}
\end{equation}

Equation~(\ref{Eq:IntegralRepresentationPsi1.7}) appears to be somewhat intriguing: the wavefield of a free-space
paraxially propagated FG beam of any order can be represented via a 1D integral defined over a {\em finite} interval.
This could seem a somewhat peculiar situation, due to the fact that the initial field distribution~[Eq.~(\ref{Eq:FlattenedGaussian.0})] has infinite
support, namely the whole plane $z=0$. But what is, in our opinion, even more important is that the integral 
representation~[Eq.~(\ref{Eq:IntegralRepresentationPsi1.7})] would hardly be reachable starting from Fresnel's integral~[Eq.~(\ref{Eq:FlattenedGaussian.0.1})],
without passing through the $\Psi_1$ function and its transformation rules. In the next section, a similar scenario will also be found as far as the more 
general problem is concerned.

\section{Paraxial propagation through $ABCD$ systems}
\label{Sec:ABCDParaxialPropagation}

\subsection{Preliminaries}
\label{Subsec:ABCDParaxialPropagationPreliminaries}

The free-space paraxial propagation formula derived in the previous section will now be extended to the  general case of the paraxial propagation
of FG beams of any order through typical paraxial optical systems with axial symmetry, characterized by the so-called $ABCD$ optical matrices.
%Although in such cases the common approach is based on Collin's integral~[???], f
For FG beams of integer order, it was found in~\cite{Borghi/2001} that the propagation problem
can be dealt with in exact terms by expanding the initial field distribution given into~Eqs.~(\ref{Eq:FlattenedGaussian.0}) and~(\ref{Eq:FGProfile}) as a 
finite superposition eLG beams as follows:
\begin{equation}
\label{Eq:FlattenedGaussianELG}
\begin{array}{l}
\displaystyle
\psi_0(\boldsymbol{r})\,=\,\sum_{n=0}^{N-1}\,(-)^n\,\left({{N}\atop{n+1}}\right)\,\mathrm{eLG}_n\left(\dfrac{\ii kr^2}{2q_N}\right)\,,
\end{array}
\end{equation}
where the symbol $\mathrm{eLG}_n(x)=\exp(x)L_n(-x)$ will be referred to as the elegant Laguerre function of order $n$
and  the complex radius of curvature $q_N=\dfrac{ka^2}{2\ii N}$ has also been introduced. 
The initial distribution $\psi_0$ is then recast as follows:
\begin{equation}
\label{Eq:FlattenedGaussianELG.1}
\begin{array}{l}
\displaystyle
\psi_0(\boldsymbol{r})\,=\,%\exp\left(\dfrac{\ii kr^2}{2q_N}\right)\,\sum_{n=0}^{N-1}\,(-)^n\,\left({{N}\atop{n+1}}\right)\,L_n\left(-\dfrac{\ii kr^2}{2q_N}\right)\,=\,
\exp\left(\dfrac{\ii kr^2}{2q_N}\right)\,\mathcal{G}_N\left(1,\,-\dfrac{\ii kr^2}{2q_N}\right)\,,%\sum_{n=0}^{N-1}\,(-)^n\,\left({{N}\atop{n+1}}\right)\,L_n\left(-\dfrac{\ii kr^2}{2q_N}\right)\,=\,\,,
\end{array}
\end{equation}
where the function $\mathcal{G}_N\left(\cdot,\cdot\right)$ is defined, for integer $N$,  as
\begin{equation}
\label{Eq:FlattenedGaussianELG.2}
\begin{array}{l}
\displaystyle
\mathcal{G}_N\left(t,\,s\right)\,=\,
\sum_{n=0}^{N-1}\,(-t)^n\,\left({{N}\atop{n+1}}\right)\,L_n(s)\,,
\end{array}
\end{equation}
In~\cite{Borghi/2001} it was proved that, if the initial field distribution given by Eq.~(\ref{Eq:FlattenedGaussianELG.1}) feeds an axially symmetric  paraxial optical system described by the optical  matrix $\mathcal{M}$
\begin{equation}
\label{Eq:FlattenedGaussianELG.0.0}
\begin{array}{l}
\displaystyle
\mathcal{M}\,=\,
\left(
\begin{array}{cc}
A & B \\
&\\
C & D
\end{array}
\right)
\,,
\end{array}
\end{equation}
then the wavefield at the output plane of  the system, say $\psi_1(\boldsymbol{r})$, takes on the following form:
\begin{equation}
\label{Eq:FlattenedGaussianELG.3}
\begin{array}{l}
\displaystyle
\psi_1(\boldsymbol{r})\,=\,\\
\\
\,=\,
\dfrac{\exp\left(\dfrac{\ii kr^2}{2Q_N}\right)}A\,
\dfrac {1}{1\,+\,\dfrac{B}{A\,q_N}}\,
\mathcal{G}_N\left(\dfrac {1}{1\,+\,\dfrac{B}{A\,q_N}},\,\dfrac {\dfrac{kr^2}{2\ii A^2\,q_N}}{1\,+\,\dfrac{B}{A\,q_N}}\right)\,,
\end{array}
\end{equation}
where an overall phase factor $\exp(\ii k \ell)$ (with $\ell$ being the optical lenght) will be tacitly assumed and $Q_N$ denotes the complex 
quantity
\begin{equation}
\label{Eq:FlattenedGaussianELG.4}
\begin{array}{l}
\displaystyle
Q_N\,=\,\dfrac{A\,q_N\,+\,B}{C\,q_N\,+\,D}\,.
\end{array}
\end{equation}
The problem of extending the function $\mathcal{G}_N\left(t,\,s\right)$ to $N \notin \mathbb{N}$ will now be addressed.

\subsection{Extension of the function $\mathcal{G}_N\left(t,\,s\right)$ to $N \notin \mathbb{N}$}
\label{Subsec:AnalyticalContinuationGN}

The starting point is the following Laplace transform representation of $\mathcal{G}_N(t,s)$ established in~\cite{Borghi/2013}:
\begin{equation}
\label{Eq:AsymptoticsLargeN.1}
\begin{array}{l}
\displaystyle
\mathcal{G}_N(t,s)\,=\,
\displaystyle
\exp(s)\,\int_0^\infty\,
\mathrm{d}\xi\,\exp (-\xi)\,
J_0\left(2\,\sqrt{s\,\xi}\right)\,
L_{N-1}^{(1)}(\xi\,{t})\,.
\end{array}
\end{equation}
%
%where $L^{(\alpha)}_{n}(\cdot)$ denotes the generalized Laguerre polynomial of order $n$ and degree $\alpha$. 
For  $N\in\mathbb{N}$, the Laguerre polynomials $L_{N-1}^{(1)}$ can be written as
\begin{equation}
\label{Eq:AsymptoticsLargeN.1.1}
\begin{array}{l}
\displaystyle
L_{N-1}^{(1)}(\xi\,{t})\,=\,
\sum_{n=0}^{N-1}\,L_{n}(\xi\,{t})\,,
\end{array}
\end{equation}
so that, on substituting from Eq.~(\ref{Eq:AsymptoticsLargeN.1.1})  into Eq.~(\ref{Eq:AsymptoticsLargeN.1}), it is found
\begin{equation}
\label{Eq:AsymptoticsLargeN.1.1.1}
\begin{array}{l}
\displaystyle
\mathcal{G}_N(t,s)\,=\,\\
\\
\,=\,
\displaystyle
\exp(s)\,
\sum_{n=0}^{N-1}
\int_0^\infty\,
\mathrm{d}\xi\,\exp (-\xi)\,
J_0\left(2\,\sqrt{s\,\xi}\right)\,
L_{n}(\xi\,{t})\,=\,\\
\\
\,=\,
\displaystyle
\sum_{n=0}^{N-1}\,
(1\,-\,t)^{n}\,L_{n}\left(\dfrac {st}{t-1}\right)\,,
\end{array}
\end{equation}
where in the last passage,~\cite[formula~3.24.6.2]{Prudnikov/Brychkov/Marichev/1986/IV} has been used.
Equation~(\ref{Eq:AsymptoticsLargeN.1.1.1}) is a valid alternative, for $N \in \mathbb{N}$, to the definition given into Eq.~(\ref{Eq:FlattenedGaussianELG.2}).
For the scopes of the present paper, its importance stems from the fact that the quantity $\mathcal{G}_N$ can also be thought of as function of two new 
variables, namely
\begin{equation}
\label{Eq:NuoveVariabili}
%\boxed
\left\{
\begin{array}{l}
\displaystyle
1\,-\,t\,=\,\dfrac 1{1\,+\,\dfrac {Aq_N}B}\,,\\
\\
\dfrac {st}{t-1}\,=\,\dfrac{\ii k r^2}{2AB}\,\dfrac 1{1\,+\,\dfrac B{Aq_N}}\,,
\end{array}
\right.
\end{equation}
and this will reveal of a certain importance in the rest of our analysis. 

In order to extend the integral into Eq.~(\ref{Eq:AsymptoticsLargeN.1}) to $N\notin \mathbb{N}$, the following notable formula, again
established by  Brychkov~\cite{Brychkov/2014}, will be employed:
\begin{equation}
\label{Eq:ClassicalFlattenedGaussianPropagatedBrychkov}
\begin{array}{l}
\displaystyle
\int_0^\infty\,x^{\alpha-1}\exp(-a x)\,J_\nu(b\sqrt x)\,L^{(\lambda)}_n(cx)\,\dd x\,=\,\\
\\
\,=\,\left(\dfrac b2\right)^\nu\,\dfrac{\Gamma\left(\alpha+\dfrac\nu 2\right)(\lambda+1)_n}{n!\,a^{\alpha+\nu/2}\Gamma(\nu+1)}\,
\Psi_1\left(
\left.
\begin{array}{l}
\alpha+\dfrac{\nu}2,-n\\
\lambda+1,\nu+1
\end{array}
\right| \dfrac ca, -\dfrac{b^2}{4a}
\right)\,.
\end{array}
\end{equation}
In particular, on letting $\alpha=1$, $a=1$, $\nu=0$, $b=2\sqrt{s}$, $t=c$, $n=N-1$, and $\lambda=1$, Laplace's transform
into Eq.~(\ref{Eq:AsymptoticsLargeN.1}) takes on the form
\begin{equation}
\label{Eq:ClassicalFlattenedGaussianPropagatedBrychkov.2}
\begin{array}{l}
\displaystyle
\mathcal{G}_N(t,s)\,=\,
N\,\exp(s)\,\Psi_1\left(
\left.
\begin{array}{l}
1,1-N\\
2,1
\end{array}
\right| t, -s
\right)\,.
\end{array}
\end{equation}
Again, it can be appreciated how the confluent hypergeometric function $\Psi_1$ constitutes the mathematical skeleton of the propagated field.
But there is more. In Appendix~\ref{App:Psi1Phi1}, the following relationship has been established:
\begin{equation}
\label{Eq:Psi1.1.1}
\begin{array}{l}
\displaystyle
\Psi_1\left(
\left.
\begin{array}{l}
1,1-N\\
2,1
\end{array}
\right| t, -s
\right)\,=\,\\
\\
\,=\,
\dfrac{\exp(-s)}{(1-t)^{1-N}}\,
\Phi_1\left(
\left.
\begin{array}{c}
1-N,1\\
2
\end{array}
\right| \dfrac t{t-1}, \dfrac{st}{t-1}
\right)\,,
\displaystyle
\end{array}
\end{equation}
where $\Phi_1$ is the confluent hypergeometric function defined by Eq.~(\ref{Eq:Phi1}).
On substituting from Eq.~(\ref{Eq:Psi1.1.1}) into Eq.~(\ref{Eq:ClassicalFlattenedGaussianPropagatedBrychkov.2}), we have
\begin{equation}
\label{Eq:Phi1.2.2}
\begin{array}{l}
\displaystyle
%\boxed{
\mathcal{G}_N(t,s)\,=\,
N\,(1-t)^{N-1}\,\Phi_1\left(
\left.
\begin{array}{c}
1-N,1\\
2
\end{array}
\right| \dfrac t{t-1}, \dfrac {st}{t-1}
\right)
%}
\end{array}
\end{equation}
so that Eq.~(\ref{Eq:FlattenedGaussianELG.3}) eventually becomes
\begin{equation}
\label{Eq:Phi1.2.2.1.1.1}
\begin{array}{l}
\displaystyle
%\boxed{
\psi_1(\boldsymbol{r})\,=\,
\exp\left(\dfrac{\ii kr^2}{2Q_N}\right)\,\dfrac{q_NN}{B}\,
\left(\dfrac {1}{1\,+\,\dfrac{A\,q_N}{B}}\right)^N\,\\
\\
\times\,
\Phi_1\left(
\left.
\begin{array}{c}
1-N,1\\
2
\end{array}
\right| 
-\dfrac{A\,q_N}{B}\,, 
\dfrac{\ii k r^2}{2AB}\,\dfrac 1{1\,+\,\dfrac B{Aq_N}}
\right)\,.
%}
\end{array}
\end{equation}
Equation~(\ref{Eq:Phi1.2.2.1.1.1})  summarizes the main result of the present paper: the general FG beam paraxial propagation problem is reduced to the evaluation of the bivariate confluent hypergeometric $\Phi_1$. 

Again, it is possible to give Eq.~(\ref{Eq:Phi1.2.2.1.1.1}) a different dress on using the following integral representation of $\Phi_1$,
established in 2012 by Brychkov and Saad~\cite[formula~3.4]{Brychkov/Saadb/2012}: % (!!!!!!!!!!!ATTENZIONE ALL'ERRORE!!!!!)
\begin{equation}
\label{Eq:Phi1.2.3.1.1}
\begin{array}{l}
\displaystyle
\Phi_1\left(
\left.
\begin{array}{c}
a,1\\
2
\end{array}
\right|  w, z
\right)\,=\,\\
\\
\,=\,
\displaystyle
(1-w)^{1-a}\,
\int_0^1\,\dd\xi\,(1-w\,\xi)^{a-2}\,{}_1F_1(a;1;z\xi)\,,
\end{array}
\end{equation}
which eventually leads to
%
%AND FINALLY ({\bf CHECKED ONLY FOR FREE PROPAGATION 17 DIC 2022})
%
%
\begin{equation}
\label{Eq:Phi1.2.2.1.1.1.Integral}
\begin{array}{l}
\displaystyle
\psi_1(\boldsymbol{r})\,=\,
\exp\left(\dfrac{\ii kr^2}{2Q_N}\right)\,\dfrac{q_NN}{B}\,
\left(\dfrac {1}{1\,+\,\dfrac{A\,q_N}{B}}\right)^N\,\\
\\
\displaystyle
\times\,
\int_0^1\,
\dfrac{\dd\xi}{\left(1\,+\,\dfrac{A\,q_N}{B}\,\xi\right)^{N+1}}\,{}_1F_1\left(1-N;1;\dfrac{\ii k r^2}{2AB}\,\dfrac \xi{1\,+\,\dfrac B{Aq_N}}\right)
\,.
\end{array}
\end{equation}
Similarly as it was found for the free-space propagation into Eq.~(\ref{Eq:IntegralRepresentationPsi1.7}), also the integral representation of 
$\psi_1$ given by Eq.~(\ref{Eq:Phi1.2.2.1.1.1.Integral}) turns out to be defined onto a finite interval $[0,1]$, despite the  infinite support of both 
the initial field distribution $\psi_0$, as well as its Fourier transform.
In the present case, however, at least a qualitative explanation of such a mathematical counterintuitive behavior can be grasped by estimating the 
right side of Eq.~(\ref{Eq:Phi1.2.2.1.1.1.Integral}) within the asymptotic limit $N\to\infty$, which corresponds to replace the initial FG 
beam distribution $\psi_0$ by that emerging from a circular hole of radius $a$. 

In particular, the asymptotics can be carried out in an elementary way, by first noting that  $Q_N\,\to\,B/D$ and that
\begin{equation}
\label{Eq:App:CircularHole.2}
\begin{array}{l}
\displaystyle
\lim_{N\to\infty}\,
\dfrac{1}{\left(1\,+\,\dfrac{A\,q_N}{B}\,\xi\right)^{N+1}}\,=\,\exp\left(\ii\dfrac{A\,ka^2}{2B}\,\xi\right)\,.
\end{array}
\end{equation}
As far as Kummer's function inside the integral is concerned, the following asymptotics holds~\cite[formula~13.8.13]{DLMF}:
\begin{equation}
\label{Eq:App:CircularHole.3}
\begin{array}{l}
\displaystyle
{}_1F_1(1-N;1;z)\,\sim\,\exp(z/2)\,J_0\left(2\sqrt{N\,z}\right)\,,\qquad N\,\gg\, 1\,,
\end{array}
\end{equation}
which, once substituted into Eq.~(\ref{Eq:Phi1.2.2.1.1.1.Integral}) together with Eq.~(\ref{Eq:App:CircularHole.2}), leads to
\begin{equation}
\label{Eq:CircularHoleBis}
\begin{array}{l}
\displaystyle
\psi_1(\boldsymbol{r})\,\sim\,
\dfrac{U}{2\ii}\,%\exp(\ii k \ell)\,
\exp\left[\ii\dfrac{UD}{2}\,\left(\dfrac ra\right)^2\right]\,\\
\\
\displaystyle
\times\,
\int_0^1\,\dd\xi\,
\exp\left(\ii \dfrac{A\,U}{2}\,\xi\right)\,J_0\left(U\,\dfrac ra\,\sqrt\xi\right)\,,\qquad\qquad N\gg 1\,,
\end{array}
\end{equation}
where now $U=ka^2/B$. 

Finally, it is not difficult to convince that Eq.~(\ref{Eq:CircularHoleBis}) is nothing but von Lommel's integral~\cite{Born/Wolf/1999}, namely, the result of 
Collins' integral for an incident  wavefield $\psi_0=\mathrm{circ}(r/a)$, as it should be expected.

\section{Conclusions}
\label{Sec:Conclusions}

Even today, the term ``superGaussian beam'' is synonymous with flat-top beam, despite the indisputable limits, both practical and theoretical, 
of the SG model and the availability of more efficient analytical approaches. For  rectangular geometries, Sedukhin's work should 
have contributed to identify flat-top profiles with an {\em error} function. 
For two-dimensional, axially symmetric geometries, Gori's and Li's models, despite allowing to solve exactly the 
paraxial propagation problem,  to date continue struggling to supplant the obsolete SG model. 

In the present paper, the FG model has been generalized to any values, no longer necessarily integer, of the order $N$. In doing this,
use has been made of the suggestion, dating back more than twenty-five years ago, by Sheppard \& Saghafi to mathematically identify the model FG through an
incomplete Gamma function. From a merely technical viewpoint, our work rests on some beautiful results recently established by Brychkov and 
co-workers. In this way, it has been possibile to analytically express the optical wave field generated by the propagation of such flat-top
``$\Gamma$-beams'' of {\em any} order through arbitrary axially symmetric paraxial optical  system (free space included)  in terms of a single bivariate 
confluent hypergeometric function. 
%We hope the present paper will help the old superGaussian model to get the deserved rest and then retire permanently. 

{As a hint for possible future works, it would be worth wondering if our results could further be generalizable to deal with the propagation of $\Gamma$-beams through nonsymmetric $ABCD$
optical systems, as well as through turbulent media. Although at present we have not yet delineated a possible development strategy, some recent extensions of the eLG superposition scheme, 
like for instance those in~\cite{Xu/Cui/Qu/2011,Xu/2013}, could be possible good inspiration sources.
}

The present model is purely analytical and provided purely analytical closed expressions of the paraxially propagated wave field.
It is a rare situation in physics in general and in optics in particular. The ubiquitous presence of less and less known special functions, such as 
bivariate hypergeometric ones certainly are, also constitutes in our opinion an added value of the present work.
We strongly encourage our readers to go through an interesting paper written more than twenty years ago by Michael Berry~\cite{Berry/2001}, 
whose content seems nowadays more than ever more relevant. In particular, the current availability of powerful computational platforms, such as 
{\em Mathematica} and {\em Maple}, will allow in the future to increase the set of special functions whose evaluation could be implemented at arbitrarily high
accuracies. We hope bivariate confluent hypergeometric functions, including of course $\Psi_1$ and $\Phi_1$, could soon become part of such a mathematical 
arsenal.

\section*{Acknowledgements}

{I am  grateful to both reviewers for their criticisms and suggestions, always aimed at improving the quality of the present work.}
I also wish to thank Turi Maria Spinozzi for his help during the preparation of the manuscript.

\appendix

\section{Proof of Eq.~(8)}
\label{App:M2}

All integrals here presented have been found with {\em Mathematica}.
The $M^2$ factor is defined by
\begin{equation}
\label{Eq.M2FactorDefinition}
\begin{array}{l}
\displaystyle
M^2\,=\,2\pi\,\sigma_r\,\sigma_p\,,
\end{array}
\end{equation}
where $\sigma_r$ and $\sigma_p$ denote the widths across the plane $z=0$ and the plane of spatial frequencies, respectively, both of them
normalized to the beam energy.
Due to the axial symmetry, $\sigma_r$ can then be expressed (in units of $a$) as follows:
\begin{equation}
\label{Eq.SigmaErre}
\begin{array}{l}
\displaystyle
\sigma^2_r\,=\,\dfrac
{\displaystyle\int_0^\infty\,\dd r\,r^3\,\psi^2_0(\boldsymbol{r})}
{\displaystyle\int_0^\infty\,\dd r\,r\,\psi^2_0(\boldsymbol{r})}\,.
%\,=\,
%\dfrac{1}{2N}\,
%\dfrac{\sqrt\pi\,\Gamma(N+2)\,-\,2{\Gamma\left(N+\dfrac 32\right)}}
%{\sqrt\pi\,\Gamma(N+1)\,-\,{\Gamma\left(N+\dfrac 12\right)}}\,.
\end{array}
\end{equation}
The denominator turns out to be 
\begin{equation}
\label{Eq.Spectral.2}
\begin{array}{l}
\displaystyle
\int_0^\infty\,\dd r\,r\,\psi^2_0(\boldsymbol{r}) %\,=\,%2\pi\,\int_0^\infty\,r\,\dd r\,\dfrac{\Gamma^2(N,N r^2)}{\Gamma^2(N)}\,=\,
%\dfrac\pi {N\,\Gamma^2(N)}\,\int_0^\infty\,\dd \xi\,\Gamma^2(N,\xi)%\,=\,%\\
%\\
\,=\,\pi\,\left[1\,-\,\dfrac{\Gamma\left(N+\dfrac 12\right)}{\sqrt\pi\,\Gamma(N+1)}\right]\,,
\end{array}
\end{equation}
while the numerator is
\begin{equation}
\label{Eq.Spectral.3}
\begin{array}{l}
\displaystyle
\int_0^\infty\,\dd r\,r^3\,\psi^2_0(\boldsymbol{r})%\,=\,%2\pi\,\int_0^\infty\,r\,\dd r\,\dfrac{\Gamma^2(N,N r^2)}{\Gamma^2(N)}\,r^2\,=\,
%\dfrac\pi {(N\,\Gamma(N))^2}\,\int_0^\infty\,\dd \xi\,\xi\,\Gamma^2(N,\xi)\,=\,
%\\
%\,=\,
%\dfrac\pi 2\,\left[1\,+\,\dfrac 1N\,-\,\dfrac{2\Gamma\left(N+\dfrac 32\right)}{\sqrt\pi\,N^2\,\Gamma(N)}\right]\,,
%\\
%\\
%\dfrac\pi 2\,\left[1\,+\,\dfrac 1N\,-\,\dfrac{2\Gamma\left(N+\dfrac 32\right)}{\sqrt\pi\,N\,\Gamma(N+1)}\right]\,,
%\\
\dfrac\pi 2\,\left[1\,+\,\dfrac 1N\,-\,\dfrac {\left(2N+1\right)}{N}\dfrac{\Gamma\left(N+\dfrac 12\right)}{\sqrt\pi\,\Gamma(N+1)}\right]\,.
\end{array}
\end{equation}

The spectral width $\sigma_p$ can also be expressed in terms of quantities defined across the plane $z=0$, being (in units of $1/a$)
\begin{equation}
\label{Eq.SigmaPi}
\begin{array}{l}
\displaystyle
\sigma^2_p\,=\,
\dfrac 1{2\pi}\,\dfrac
{\displaystyle
\int_0^\infty\,\dd r\,r\,\left(\dfrac{\partial\psi_0}{\partial r}\right)^2}
{\displaystyle
\int_0^\infty\,\dd r\,r\,\psi^2_0(\boldsymbol{r})}\,,
\end{array}
\end{equation}
where the numerator turns out to be
\begin{equation}
\label{Eq.SigmaPi.1}
\begin{array}{l}
\displaystyle
{\displaystyle
\int_0^\infty\,\dd r\,r\,\left(\dfrac{\partial\psi_0}{\partial r}\right)^2}\,=\,2^{1-2N}\,\Gamma(2N)\,,
\end{array}
\end{equation}
so that, on using again Eq.~(\ref{Eq:DerivativeGamma}), 
\begin{equation}
\label{Eq.SigmaPiNew}
\begin{array}{l}
\displaystyle
\sigma^2_p\,=\,
\dfrac{1}{\pi^2\,2^{2N}\,\Gamma(N)^2}\,
\dfrac{\sqrt\pi\,\Gamma(N+2)\,{\Gamma(2N)}}
{\sqrt\pi\,\Gamma(N+1)\,-\,{\Gamma\left(N+\dfrac 12\right)}}\,.
\end{array}
\end{equation}
Finally, on substituting from Eqs.~(\ref{Eq.SigmaErre})~and~(\ref{Eq.SigmaPiNew}) into Eq.~(\ref{Eq.M2FactorDefinition}), Eq.~(\ref{Eq.M2FactorFG}) follows.

\section{Solving Eq.~(12)}
\label{App:Cardano}

{
Consider the following fourth order algebraic equation:
\begin{equation}
\label{Eq:Cardano.1}
\begin{array}{l}
\displaystyle
X^4\,-\,B\,X^3\,+\,A\,=\,0\,,
\end{array}
\end{equation}
with $A$ and $B$ real positive parameters. 
On using the {\em Mathematica} implementation of Cardano's formula, it is not difficult to prove that the real root we are interested in can be expressed through the following algorithm:
\begin{equation}
\label{Eq:Cardano.2}
\begin{array}{l}
\displaystyle
X\,=\,
\dfrac B4 
\left[1 \,+\, \sqrt{1 + \dfrac {4\Xi}{B^2}} 
\,+\,
\sqrt{
   2 \left(
   1 \,+\, \dfrac 1{\sqrt{1 + \dfrac {4\Xi}{B^2}}}
   \right) \,-\, \dfrac {4\Xi}{B^2}}
  \right]\,,
\end{array}
\end{equation}
where
\begin{equation}
\label{Eq:Cardano.3}
\begin{array}{l}
\displaystyle
\Xi\,=\,
\dfrac {4A\,\left(\dfrac 23\right)^{1/3}}{\Delta}\,+\,\dfrac {\Delta}{18^{1/3}}\,,
\end{array}
\end{equation}
and
\begin{equation}
\label{Eq:Cardano.4}
\begin{array}{l}
\displaystyle
\Delta\,=\,\left[9AB^2\,+\,\left(1\,+\,\sqrt{1\,-\,\dfrac{256}{27}\,\dfrac A{B^4}}\right)\right]^{1/3}\,.
\end{array}
\end{equation}
}

\section{Proof of Eq.~(23)}
\label{App:Choi/Hasanov}

Due to the 2011 paper by Choi and Hasanov~\cite{Choi/Hasanov/2011}, the following integral representation of $\Psi_1$ can be established:
\begin{equation}
\label{Eq:IntegralRepresentationPsi1}
\begin{array}{l}
\displaystyle
\Psi_1\left(
\left.
\begin{array}{l}
N+1,1\\
N+1,1
\end{array}
\right| x, y
\right)\,=\,\dfrac{\Gamma(\epsilon)}{\Gamma(N)\Gamma(\epsilon-N-1)}\,\times\\
\\
\displaystyle
\int_0^1\int_0^1\,\dd\xi\,\dd\eta\,
\dfrac{\eta^N(1-\xi)^{N-1}(1-\eta)^{\epsilon-N-2}}{(1-x\xi)^{N+1}}\,\\
\\
\displaystyle
\times\,
\exp\left(-\dfrac {y\eta}{x\xi-1}\right)\,{}_1F_1\left(1-\epsilon;1;\dfrac {y\eta}{x\xi-1}\right)
\end{array}
\end{equation}
where $\epsilon$ denotes an arbitrary complex parameters which must only satisfy the condition $\mathrm{Re}\{\epsilon\} > \mathrm{Re}\{N\} +1$. 
In particular, on letting $\epsilon=N+2$, Eq.~(\ref{Eq:IntegralRepresentationPsi1}) yields
\begin{equation}
\label{Eq:IntegralRepresentationPsi1.2}
\begin{array}{l}
\displaystyle
\Psi_1\left(
\left.
\begin{array}{l}
N+1,1\\
N+1,1
\end{array}
\right| x, y
\right)
\,=\,\dfrac{\Gamma(N+2)}{\Gamma(N)\Gamma(1)}\,\times\\
\\
\displaystyle
\int_0^1\dd\xi\,\dfrac{(1-\xi)^{N-1}}{(1-x\xi)^{N+1}}\,\,\\
\\
\displaystyle
\times\,
\int_0^1\,\dd\eta\,
\eta^N\,
\exp\left(-\dfrac {y\eta}{x\xi-1}\right)\,{}_1F_1\left(-N-1;1;\dfrac {y\eta}{x\xi-1}\right)\,=\,\\
\\
\,=\,\dfrac{\Gamma(N+2)}{\Gamma(N)}\,\,\\
\\
\displaystyle
\times\,
\displaystyle
\int_0^1\dd\xi\,\dfrac{(1-\xi)^{N-1}}{(1-x\xi)^{N+1}}\,\int_0^1\,\dd\eta\,
\eta^N\,
%\exp\left(-\dfrac {y\eta}{x\xi-1}\right)\,
{}_1F_1\left(N+2;1;\dfrac {y\eta}{1-x\xi}\right)%\,=\,\\
\,,
\end{array}
\end{equation}
where, in the last step, Kummer's transformation has been employed. 
The inner $\eta$ integral can be evaluated by using~\cite[formula~2.21.1.4]{Prudnikov/Brychkov/Marichev/1986/III}, which yields 
\begin{equation}
\label{Eq:PrudnikovII.2.19.3.6.1}
\begin{array}{l}
\displaystyle
\int_0^1\,\dd\eta\,\eta^N\,{}_1F_1\left(N+2;1;\dfrac {y\eta}{1-x\xi}\right)\,=\,\,\\
\\
\displaystyle
\,=\,
\dfrac 1{N+1}\,
{}_1F_1\left(N+1;1;\dfrac {y}{1-x\xi}\right)\,.
\end{array}
\end{equation}
Finally, on substituting from Eq.~(\ref{Eq:PrudnikovII.2.19.3.6.1}) into Eq.~(\ref{Eq:IntegralRepresentationPsi1.2}), after simple  algebra
Eq.~(\ref{Eq:IntegralRepresentationPsi1.7}) follows.

\section{Proof of Eq.~(36)}
\label{App:Psi1Phi1}

From the very definition into Eq.~(\ref{Eq:Psi1}) we have
\begin{equation}
\label{Eq:Psi1.1}
\begin{array}{l}
\displaystyle
\Psi_1\left(
\left.
\begin{array}{l}
1,\beta\\
2,1
\end{array}
\right| t, -s
\right)\,=\,
\sum_{k=0}^\infty\,\sum_{\ell=0}^\infty\,\dfrac{(1)_{k+\ell}\,(\beta)_k}{(2)_k(1)_\ell}\,\dfrac{t^k}{k!}\,\dfrac{(-s)^l}{\ell!}\,=\,\\
\\
\,=\,
\displaystyle
\sum_{k=0}^\infty\,\dfrac{(1)_{k}\,(\beta)_k}{(2)_k}\,\dfrac{t^k}{k!}
\sum_{\ell=0}^\infty\,\dfrac{(1+k)_{\ell}}{(1)_\ell}\,\,\dfrac{(-s)^l}{\ell!}\,=\,\\
\\
\,=\,
\displaystyle
\sum_{k=0}^\infty\,\dfrac{(1)_{k}\,(\beta)_k}{(2)_k}\,\dfrac{t^k}{k!}
{}_1F_1(1+k;1;-s)\,=\,\\
\\
\,=\,
\displaystyle
\exp(-s)\,\sum_{k=0}^\infty\,\dfrac{(\beta)_k}{(2)_k}\,t^k
L_k(s)\,.
\end{array}
\end{equation}
Last series can be expressed in closed form via~\cite[5.11.2.7]{Prudnikov/Brychkov/Marichev/1986}, i.e.,
\begin{equation}
\label{PrudnikovII.5.11.2.7}
\begin{array}{l}
\displaystyle
\sum_{k=0}^\infty\,\dfrac{(a)_k\,t^k}{(\alpha+\beta)_k}\,L^\alpha_k(x)=
(1-t)^{-a}
\Phi_1\left(
\left.
\begin{array}{l}
a,\beta-1\\
\alpha+\beta
\end{array}
\right| \dfrac t{t-1}, \dfrac{tx}{t-1}
\right),
\end{array}
\end{equation}
from which, on letting $a=\beta$, $\alpha=0$, $\beta=2$, and $x=s$, 
after straightforward algebra  Eq.~(\ref{Eq:Psi1.1.1}) follows.

%%%%%%%%%%%%%%%%%%%%%%% References %%%%%%%%%%%%%%%%%%%%%%%%%


\begin{thebibliography}{00}
\bibitem{FootNote} Equation~(\ref{Eq:FGProfile}) was originally derived starting from the identity $1\,=\,\exp(-\xi^2)\,\exp(\xi^2)$ and on truncating the Taylor expansion of the second exponential up to $N$.
In the present paper, however, we restrict the expansion to the first $N$ terms. With such choice the case $N=1$ correspond to the Gaussian beam.
\end{thebibliography}
\end{document}